\documentclass[
aps,%
12pt,%
final,%
notitlepage,%
oneside,%
onecolumn,%
nobibnotes,%
nofootinbib,%
superscriptaddress,%
showpacs,%
centertags]%
{revtex4}
\def\be {\begin {eqnarray}}
\def\ee {\end {eqnarray}}

\newcommand{\bea}{\begin{eqnarray}}
\newcommand{\eea}{\end{eqnarray}}

\begin{document}
\title
{Is it possible to estimate the Higgs Mass from \\ the CMB Power Spectrum?
}

\author{\firstname{A.B.}~\surname{Arbuzov}}
\affiliation{Joint Institute for Nuclear Research, Dubna, Russia}
\author{\firstname{B.M.}~\surname{Barbashov}}
\affiliation{Joint Institute for Nuclear Research, Dubna, Russia}
\author{\firstname{A.}~\surname{ Borowiec}}
\affiliation{Institute of Theoretical Physics, University of
 Wroc\l aw, Wroc\l aw, Poland}
\author{\firstname{V.N.}~\surname{Pervushin}}
\affiliation{Joint Institute for Nuclear Research, Dubna, Russia}
\author{\firstname{S.A.}~\surname{Shuvalov}}
\affiliation{Russian Peoples Friendship
University, Moscow, Russia}
\author{\firstname{A.F.}~\surname{Zakharov}}
\affiliation{Institute of Theoretical and
Experimental Physics, Moscow,
Russia}

\begin{abstract}
General Relativity  and Standard Model are considered as a theory of dynamical scale symmetry
 with definite initial data compatible with the accepted Higgs mechanism.
 In this theory
the Early Universe behaves like a factory of
electroweak bosons and Higgs scalars, and it gives a possibility to identify three peaks in the Cosmic Microwave
Background power spectrum with the contributions of photonic decays
and annihilation processes of primordial Higgs, $W$, and $Z$  bosons
in agreement with the QED coupling constant,   Weinberg's angle, and
    Higgs' particle mass of about $118$ GeV.
\end{abstract}

\maketitle
\section{Introduction}\label{int1}

The observational data~\cite{WMAP01} on the  Cosmic Microwave Background (CMB)  power spectrum
show several clear peaks at the orbital momenta $\ell_1\simeq 220$,
$\ell_2\simeq 546$, $\ell_3\simeq 800$. These phenomena are explained in the
$\Lambda$CDM model~\cite{Giovannini} by acoustic inhomogeneities of
the scalar metric component
 treated as a dynamical variable. 
 By adjusting parameters of the equations for the acoustic
excitations one can provide a good fit of the observed peaks and predict
other peaks with higher $\ell$ values, which can be found in future observations.
Recall that  the $\Lambda$CDM  model requires  the acoustic
explanation of the CMB power spectrum~ by introduction of a
dynamical scalar metric component that is absent in the Wigner  classification of relativistic
states~\cite{Wigner_39}. The dynamical
scalar metric component is introduced by the $\Lambda$CDM model
 without any substantial motivation and clear discussion
of the reasons for introducing new concepts. Moreover, this $\Lambda$CDM explanation contradicts to
the vacuum postulate.
Since the  CMB power spectrum is one of the highlights of the
present-day Cosmology with far-reaching implications and more
precise observations are planned for near future \cite{WMAP01},
the detailed investigation of any possible flaw of the standard
theory deserves an attention and a public discussion.

In this paper we try to describe  the CMB power spectrum in accord with the well-established
Wigner's theory of the relativistic state classification, where   any relativistic particle in quantum
field theory can be associated with a unitary irreducible representation of
the Poincar\'e group given in a definite frame  with a  positive energy.

 The  cosmological \textit{scale} factor, its local excitations
used  for description of the CMB power spectrum, and Poincar\'e group transformations
can be naturally included in the Wigner classification, if General Relativity is considered as
the theory   of  the joint non-linear realization of the affine and conformal symmetries with the
Poincar\'e group of the vacuum stability~\cite{og}, where the \textit{scale} invariance of laws of
Nature \cite{Weyl_18,Dirac_73}
is realized dynamically by means of  the dilaton Goldstone field.

The dynamical scale symmetry plays a role of
 the principle of a choice of variables in the accepted General Relativity (GR)
and the Standard Model (SM) \cite{1998}.
 The dilaton Goldstone field  compensates all scale transformations of  fields
including the cosmological scale factor describing expansion of the Universe lengths in
the Standard Cosmology~\cite{Giovannini}. Nevertheless,
the cosmological dynamics can be introduced by help of Einstein's cosmological principle \cite{Einstein_17}
that means averaging all scalar characteristics including the dilaton field
over a constant Universe volume. This cosmological dynamics of the zeroth dilaton mode
explains the redshift by a permanent increase of all masses in the Universe and leads
to  the Conformal Cosmology~\cite{Behnke_02,Behnke_04,zakhy,Blaschke_04,Barbashov_06,B_06,Pervushin_07,Glinka_08},
where all measurable quantities are identified with the conformal ones
(conformal time, coordinate distance, and constant conformal temperature).

General Relativity considered as the theory dynamical scale symmetry~\cite{og,Dirac_73}
changes the numerical analysis of
supernovae type Ia data~\cite{Riess_98,Riess_04}
and  shows the dominance of  the scalar field \textit{kinetic} energy
in all epochs of the Universe evolution including the chemical evolution, recombination,
and SN explosions.


In the paper we try to describe the CMB power spectrum \cite{WMAP08} in GR as the theory  of dynamical
scale symmetry
in accord with the classification of relativistic states \cite{Wigner_39}.

 \section{Dilatonic variables in General Relativity}

Let us consider the accepted  General Relativity  supplemented by  the  Standard Model
and an additional scalar field $Q$ governing the Universe evolution
 \be\label{1-1}
\frac{}{} S_{\rm U}[g,F]=\int d^4x\sqrt{-g}\left[-\frac{R(g)}{6}
 +{\cal L}_{\rm SM}(F)+\partial_\mu Q\partial^\mu Q \right],
 \ee
where units
 $\hbar=c=M_{\rm Planck}\sqrt{{3}/({8\pi}})=1$
are used throughout the paper.
This action depends on a set of scalar, spinor, vector, and  tensor fields
$F_{(n)}=\phi,s,V_{\mu},g_{\mu\nu}$
with their conformal weights $n=-1,-3/2,0,2$, respectively.

Following the foundation of the GR as a dynamical scale symmetry~\cite{Dirac_73,og}
we define all observable fields $\widetilde{F}_{(n)}$ as scale-invariants quantities
using the following scale transformations of  these fields $F_{(n)}$ in action (\ref{1-1})
including the metric components $g_{\mu\nu}$:
\be\label{1-2}
\widetilde{F}_{(n)}=\exp\{nD\}F_{(n)},\qquad \widetilde{g}_{\mu\nu}=\exp\{2D\}g_{\mu\nu},
\ee
where $D$ is the dilaton  compensating scale transformations of all these fields.
Any concrete choice of the dilaton as a metric functional $D[g]$ means a gauge fixing.
In \cite{dir,lich,Blaschke_04} this functional is chosen in the form of $D[g]=-\log |g^{(3)}|/6$ in
according with the accepted definition
of transverse and traceless graviton physical variables
 given in a definite frame distinguishing the spatial metric components $g^{(3)}_{ij}$.
Therefore, one can remove any scale factor from the spatial metric components in the Dirac-ADM
parameterization~\cite{dir} in terms of the simplex components
$\widetilde{\omega}_{(0)}$, $\widetilde{\omega}_{(b)}$ in the Minkowskian tangent space-time
 \bea \label{dg-2}
 \widetilde{ds}^2=\widetilde{\omega}^2_{(0)}-\widetilde{\omega}^2_{(b)},\quad
 \widetilde{\omega}_{(0)}=e^{-2{D}}{N}_{\rm d}dx^0,\quad
 \widetilde{\omega}_{(b)}={\bf e}_{(b)j}(dx^j+N^j dx^0),
 \eea
where ${\bf e}_{(b)i}$ are the triads~\cite{lich} with the unit spatial metric
determinant $|{\bf e}_{(b)i}{\bf e}_{(b)j}|=1$, ${N}_{\rm d}$ is the Dirac lapse function,
and $N^j$ are the shift vector components. In phenomenological applications,
one can identify this choice with the CMB co-moving reference frame.
In terms of the dilaton variables, the GR action takes the form
 \bea\label{1-3dg}
 {S}_{\rm GR}=
 -\!\!\int\limits_{ }^{ }d^4x\sqrt{-g}\frac{R(g)}{6}\!\!=\!\!\int\limits_{}^{} d^4x
 \left[-{\frac{v^2_{{D}}}{N_{\rm d}}}+\frac{v^2_{(ab)}}{24N_{\rm d}}-N_{\rm d}e^{-4D}\frac{ R^{(3)}({\bf e})+8e^{D/2}\triangle e^{-D/2}}{6}\right],
 \eea
where $R^{(3)}({\bf{e}})$ is a curvature,
$\triangle =\partial_i[{\bf e}^i_{(a)}{\bf e}^j_{(a)}\partial_j]$ is the  Laplace operator,
and
 $v_{{D}}=\left[\check{\partial}_0{D} +\partial_lN^l/3\right]$,
 $v_{(ab)}={\bf e}_{(a)i}v^i_{(b)}+{\bf e}_{(b)i}v^i_{(a)},\quad
 v_{(a)i}= [{\check{\partial}_0{\bf e}_{(a)i}}
 + {{\bf e}_{(a)i}\partial_lN^l-{\bf e}_{(a)l}\partial_iN^l}/{3}],$
are velocities of the metric components, $\check{\partial}_0=(\partial_0-N^l\partial_l)$.

 Simplex (\ref{dg-2}) as an object of frame transformations from the Earth frame to the CMB one
 moving to Leo with the measurable velocity 368 km/s
 separates the latter from the unmeasurable
 diffeomorphisms $x^0\to \widetilde{x}^0=\widetilde{x}^0(x^0)$,
 $x^k\to \widetilde{x}^k=\widetilde{x}^0(x^0,x^k)$.
 The principle of diffeo(d)-invariance of observables, $D(x^0)=D(\widetilde{x}^0)$,
 is at heart of GR.  One can see that variables (\ref{1-2}) and interval (\ref{dg-2})
define a d-invariant finite coordinate volume $\int_{V_0} d^3x=V_0<\infty$,
 d-invariant evolution parameter in the field space of events, and a d-invariant time-interval $N_0dx^0=d\tau$
by Einstein's cosmological principle \cite{Einstein_17} as
averaging of the dilaton $D$ and the inverse Dirac lapse function $N^{-1}_{\rm d}$ over this volume:
  \be\label{1-3}
  V_0^{-1}\int_{V_0} d^3x D(\tau,x^k)=\langle D\rangle(\tau),~~~~~~~V_0^{-1}\int_{V_0} d^3x N^{-1}_{\rm d}=\langle N^{-1}_{\rm d}\rangle=N^{-1}_0.
  \ee
The scale-invariant variables and
 d-invariant evolution parameter as  the dilaton zeroth mode $\langle D\rangle (\tau)$
are compatible with a definite d-invariant cosmological dynamics known as
the Conformal Cosmology~\cite{Behnke_02,Behnke_04,zakhy,Blaschke_04,Barbashov_06,B_06,Pervushin_07,Glinka_08}
that strongly differs from the heuristic phenomenology of the accepted Standard Cosmology
\cite{Giovannini}.
Principles of the conformal symmetry, relativistic (frame) symmetry, and  d-invariance of observables
 and the Dirac Hamiltonian approach to GR
completely determine the finite volume generalization of Einstein theory ~\cite{Barbashov_06,B_06,Pervushin_07,Glinka_08}
 \bea\label{1-1dg}
 S_{\rm U}[D,\widetilde{F}]\Big |_{D=\langle  D\rangle+\overline{D}}&=&S_{\rm z}[\langle  D\rangle]
 + \widetilde{S}_{\rm U}[D,\widetilde{F}],
 \eea
where
\bea\label{1-1dg-a}
 S_{\rm z}[\langle  D\rangle]&=&
 V_0\int\limits_{\tau=0}^{\tau_0} d\tau [-(\partial_\tau\langle  D\rangle)^2+(\partial_\tau\langle \phi \rangle)^2+(\partial_\tau\langle  Q \rangle)^2]\Big|_{\,{\Large d\tau=N_0dx^0}}
 \eea
 is the zeroth mode action and  the second term $\widetilde{S}_{\rm U}$ repeats  actions (\ref{1-1}) and (\ref{1-3dg})  for nonzero harmonics  associated with
    local excitations.

\section{Cosmological dynamics of the zeroth dilaton mode}

  Let us consider the Early Universe when one can neglect all these local excitations $\widetilde{S}_{\rm U}\simeq0$ (complete expressions of  action (\ref{1-1dg}) see in Appendix A). In this case,
 the  cosmological evolution of the Empty Universe arises  in the form of a conformal
   mechanics of zeroth harmonics of all scalar fields $F=D,\phi,Q$
   with equations  $\partial^2_{\tau}\langle F\rangle=0$
    and the initial data
\be\label{id-2a}\langle \phi \rangle_I=M_W/(g\sqrt{2}),~~~ \partial_{\tau}{\langle \phi\rangle_I}=0; ~~~\langle Q\rangle_I=Q_0,~~~~~
\partial_{\tau}{\langle Q\rangle_I}=H_0
\ee
defined so that the mechanism of spontaneous  electroweak symmetry breaking does not
differ from the accepted one in SM. Recall, that the accepted spontaneous symmetry breaking
mechanism is based on   the  Coleman--Weinberg potential equation
$d\textbf{V}(\langle\phi\rangle_I)/d\langle\phi\rangle_I=0$ in the perturbation theory
restricted by the constraint $\partial_{\tau}\langle \phi\rangle=0$.
In our  perturbation theory (\ref{id-2a}), loop diagrams
also lead to  the effective potential with the same equation $d\textbf{V}(\langle\phi\rangle_I)/d\langle\phi\rangle_I=0$
treated as a constraint that
keeps the vacuum equation $\partial^2_{\tau}{\langle \phi\rangle}=0$.

Using as example this potential free model of the Empty Universe, one can see that
 the Standard Cosmology observable quantities are connected with
the conformal ones by relation~(\ref{1-2}):
  \be \label{cc-sc}F_{(n)\rm SC}=e^{-n\langle D\rangle}\widetilde{F}_{(n)\rm CC}.
  \ee
This relation determines the scale factor
\be \label{cc-sc-1}e^{-\langle D\rangle}=a(z)=(1+z)^{-1},
  \ee
conformal masses and time
\be
 \label{cc-sc-2}\widetilde{m}=a(z) m_0, \qquad d\eta=d\tau a^2(\tau),
\ee
and the horizon $\widetilde{H}=H_0 a^{-2}$.
In this case, the dilaton  solution of the motion equation $\partial^2_\tau \left\langle D \right\rangle=0$
takes the form
\be \label{cc-sc-5}
\left\langle D \right\rangle  = \left\langle D \right\rangle _0  + H_0(\tau  - \tau _0 ).
  \ee
 In terms of the effective cosmological factor~(\ref{cc-sc-1}) and conformal time~(\ref{cc-sc-2})
 this solution becomes
 \be \label{cc-sc-6}
a(\eta )= a_0 \sqrt{1  + 2H_0 (\eta  - \eta _0 )}.
  \ee
The cosmological dynamics of the Conformal Cosmology (CC)
strongly differs from the heuristic phenomenology of
the accepted  Standard Cosmology (SC) including the $\Lambda CDM$ model~\cite{Giovannini}
by a constant measurable volume
 defined by Eqs. (\ref{dg-2}), running masses, conformal time (\ref{cc-sc-2}),
 and the constant CMB conformal
temperature $T_{CC}=T_{SC} a(z)=2.725$~K during the cosmological
evolution process.
The dilaton variables (\ref{1-2}) and (\ref{dg-2})  explain redshift by
the permanent increase of all masses in the Universe
~\cite{Behnke_02,Behnke_04,zakhy,Blaschke_04,Barbashov_06,B_06,Glinka_08}.
The corresponding luminosity-distance -- redshift relation
$H_0\widetilde{\ell}(z)=z+{z^2}/{2}$ ~\cite{Behnke_02} does not contradict the recent SN
data~\cite{Riess_98} analyzed in the framework of the Conformal
Cosmology~\cite{Behnke_02,zakhy}, where the redshift is explained by running masses~(\ref{cc-sc-2}),
in the case of the conformal mechanics (\ref{id-2a}) leading to  the  rigid state
dominance $H_0\gg  \langle\sqrt{T_{\rm d}}\rangle$.

Calculation of the primordial helium abundance \cite{Wein_73,Behnke_04} takes into account
  weak interactions,
 the Boltzmann factor, (n/p) $ e^{\triangle m/T} \sim 1/6$, where $\triangle m$ is the neutron-proton
mass difference, which is the same for both SC and CC,
$\triangle m_{SC}/T_{SC}=\triangle m_{CC}/T_{CC}=(1+z)^{-1}m_{0}/T_{0}$,
and the square root dependence of the z-factor on the measurable time-interval defined in Eqs.~(\ref{cc-sc-2}) and (\ref{cc-sc-6})
  $(1+z)^{-1}\simeq \sqrt{1+2H_0(\eta-\eta_0)}$ 
explained by the dominant rigid state.
In SC, where the measurable time-interval is identified with
the Friedmann time, this square root dependence of the z-factor is explained by the
radiation dominance. 

Quantization of the theory constructs a vacuum state with minimal energy
defined as $\textsf{E}_{\rm U}=P_{\langle D\rangle}=2V_0\partial_\tau \langle D\rangle$
with the conservation law $\partial_{\langle D\rangle}\textsf{E}_{\rm U}=0$
\cite{Barbashov_06,B_06,Pervushin_07,Glinka_08}
(see Appendix B).

\section{The Early Universe as Factory of  Higgs Particles}

It was shown in Ref.~\cite{Blaschke_04} that the Empty Universe acts as a factory
of longitudinal vector bosons and Higgs particles ($h$) distinguished by their
direct interaction with the dilaton.
In particular, the field equations for creation and annihilation operators take the form\\
  \centerline{$\partial_\eta \widetilde{F}^{\pm}(\textbf{k},\eta)= \pm i\sqrt{{\bf k}^2+\widetilde{m}^2_{F}}\,\,\widetilde{F}^{\pm}(\textbf{k},\eta)
    +i \partial_\eta \langle D\rangle(\eta)
   \widetilde{F}^{\mp}(\textbf{k},\eta)+i [\textsf{H}_{\rm int}, \widetilde{F}^{\pm}(\textbf{k},\eta)]
   $.}\\
 The third  term   leads to collisions and  the Boltzmann-type distribution~\cite{Smolyansky_02}
 \be\label{kle-1}
B({\bf k},\widetilde{T}_{\rm F})=\left\{
\exp\left[\left({\sqrt{{\bf k}^2+\widetilde{m}^2_{F}}- \widetilde{m}_{F}}\right)/{ k_{\rm B}\widetilde{T}_{\rm F}}\right]
-1\right\}^{-1}.
 \ee
Here the conformal boson temperature $\widetilde{T}_{\rm F}\sim T_0$ is determined by
the collision integral kinetic equation
$
  \widetilde{n}(\widetilde{T})=[\widetilde{\sigma}_{\rm F~ scat}r_{\rm F}]^{-1} 
$, 
where   $\widetilde{n}(\widetilde{T}_{\rm F})$ is the particle number density and
$\sigma_{\rm F~ scat}$ is the cross section,  if  the free length $r_{\rm F}$
is identified with  the horizon $\widetilde{d}(z)=a(z)^2H^{-1}_0$
in CC~\cite{Blaschke_04,ber-85}
 \be\label{1-3-w}
 r_{\rm F}=[\widetilde{n}(\widetilde{T}_{\rm F})\widetilde{\sigma}_{\rm F~ scat}]^{-1}
 \simeq \widetilde{d}(z)=a(z)^2H^{-1}_0.
 \ee
Creation of these primordial particles started at the moment $a_{W\rm I}$
when their wavelengths coincided with the horizon length
$  \widetilde{M}^{-1}_{ W \rm I} =[a_{W\rm I} M_{W 0}]^{-1}\sim
\widetilde{H}_{W\rm I}^{-1}=a^2_{W\rm I} (H_{0})^{-1}$,
as it follows from the uncertainty  principle. This gives
the instance of creation of  primordial particles, in particular, $W$-bosons
 \be\label{cr-3}
 a^3_{W\rm I}\sim \frac{H_{0}}{M_{W 0}}\simeq 19 \cdot 10^{-45}
 ~\to ~ a_{W\rm I}\simeq  2.7\cdot 10^{-15}.
 \ee
The conformal photon temperature value $\widetilde{T}_\gamma=T_\gamma(z) a(z)=T_\gamma(0)$
can be estimated  from  the kinetic equation (\ref{1-3-w}).
If  $\widetilde{n}(\widetilde{T}_{\rm F})\sim  \widetilde{T}_\gamma^3$,
one can see that this temperature value
 $
 T_\gamma (0)\approx (\widetilde{M}_{WI}^2\widetilde{H}_{W\rm I})^{1/3}=(M_{\rm W0}^2H_0)^{1/3}=2.3 K
 $ 
is astonishingly close to the observed temperature $T_0=2.725$~K of the
cosmic microwave background radiation. The latter can be treated as the final decay product
of the primordial bosons that inherits their temperature.

The lifetime $\eta_L$ of  primordial bosons in the early Universe  can be estimated
by using the equation of state $a^2_{W\rm L}=a^2(\eta_L)=a_{W\rm I}^2[1+2\widetilde{H}_{W\rm I}(\eta_L-\eta_I)]$ and the
$W$-boson lifetime within the Standard Model. Specifically, we have
\be \label{life} \frac{a^2_{W\rm L}}{a^2_{W\rm I}}= 1+ 2\widetilde{H}_{W\rm I}(\eta_L-\eta_I) \simeq \frac{\widetilde{H}_{W\rm I}}{\widetilde{M}_{WL}}\frac{2\sin^2\theta_{(W)}}
{\alpha}= \frac{a_{W\rm I}}{a_{W\rm L}}\,\frac{2\sin^2
\theta_{(W)}}{\alpha},
\ee
where $\theta_{(W)}$ is the Weinberg angle,  $\alpha
=1/137$, and $
\widetilde{M}_{{W\rm I}}\simeq \widetilde{H}_{W\rm I}$.
From the solution of Eq.~(\ref{life}), $ a_{W\rm L}/a_{W\rm I}=
\left({2\sin^2\theta_{(W)}}/{\alpha
}\right)^{2/3} \simeq {16}$ it
follows that the lifetime of primordial bosons
is an order of magnitude longer than the Universe relaxation time $\eta_I=(2\widetilde{H}_{W\rm I})^{-1}$:
\be \label{lv} \eta_L-\eta_I \simeq
15(2\widetilde{H}_{W\rm I})^{-1}. \ee

The problem is to obtain parameters of the diffusion reaction system
arising in this case from the Standard Model
computing the relevant cross sections and decay rates.

 The present-day data of photon density inherits the primordial vector boson density,
 $\Omega _{\rm rad}\simeq M_{W0}^2 \cdot a_{WI}^{-2}$ $= 10^{-34}10^{29} \sim 10^{-5}$~\cite{Blaschke_04}.

In the same way intensive creation of vector bosons in the Early Universe
leads to ${\rm CP}$  non-conservation~\cite{Blaschke_04} due to the ABJ anomaly
and the CKM-mixing in the fermion--$W$-boson interaction.
So that the cosmological evolution and this  non-conservation freezes the fermion number.
This leads to the baryon-number density $n_b=n_\gamma X_{\rm CP}$,
where the factor $X_{\rm CP}\sim 10^{-9}$ is determined by the superweak
interaction of $d$ and $s$ quarks, which
is responsible for CP violation experimentally observed in
$K$-meson decays~\cite{ufn} (see Appendix C).

  The present-day baryon density $\Omega _{\rm b}\simeq 2 \alpha_W \simeq 0.06$ is calculated by the evolution of the baryon density
  from the early stage, when it was directly related to the photon density.

\section{The CMB temperature anisotropy}
The collision integral equation (\ref{1-3-w}) generalized to anisotropic
decays $T_0\to T_0+\triangle T$, $\sigma\to \sigma+\sigma_{\gamma\gamma}$ gives us
the formula
 $\big|{\triangle T}/{T_0}\big|\simeq
 ({2}/{3})\big|{\sigma_{\gamma\gamma}}/{\sigma}\big|$
and a possibility to estimate the magnitude of the CMB anisotropy.
Its observational value  about $10^{-5}\sim \alpha^2$  \cite{WMAP08}
testifies  to the dominance of the two photon processes. Therefore,
the CMB anisotropy revealed  in the region of the
three peaks $\ell_1\simeq 220$,  $\ell_2\simeq 546$, and $\ell_3\simeq800$
can reflect parameters of the primordial bosons and their
decay processes, in particular $h\to \gamma\gamma$, $W^+W^-\to
\gamma\gamma$, and $ZZ \to \gamma\gamma$.
The values of multipole momenta at the peaks can be obtained by a simple
dimensional analysis using the accepted formula~\cite{hu01}
 \be \label{cr-6a}
 \ell_{Pd}= \widetilde{d}_{Pd} \widetilde{M}_{P d}={d}_{Pd} {M }_{P 0},
 \ee
where $\widetilde{d}_{Pd}=a_{Pd}^2H_0^{-1}=a_{Pd}^{-1}{d}_{Pd}$ is the conformal
horizon  (\ref{1-3-w}) at the instances of the processes $(P)$
$h\to \gamma\gamma$, $W^+W^-\to\gamma\gamma$, and $ZZ \to \gamma\gamma$
marked by the corresponding cosmological scale factor $a_{Pd}$,
and $\widetilde{M}_{P d}=M_{P0}a_{Pd}$ is the
conformal mass of emitters in the given process.
One can see that the more horizon-length the more number of emitters covered by the horizon,
and the  more values of multipole momenta.
The substitutions of $\widetilde{d}_{Pd}=a_{Pd}^2H_0^{-1}$ and
$\widetilde{M}_{P d}=M_{P0}a_{Pd}$ into the Eq.~(\ref{cr-6a})  give us
the multipole momenta
  \be\label{ci-8}
  \ell_{Pd}=\frac{a_{Pd}^3}{H_0} M_{P0}=\frac{a_{Pd}^3}{a_{PI}^3},~~
  \ee
where  the initial data  $M_{P0}/H_0=1/a_{PI}^3$ given by Eq. (\ref{cr-3}) is taken into account,
and $\ell_{hd}=\ell_{1},~\ell_{Wd}=\ell_{2},~\ell_{Zd}=\ell_{3}$.

Identifying all photon energies  in these processes
with the mean photon one $k_{\rm eff }$ in the CMB,  multiplied by the
corresponding $z$-factor $a_{Pd}/a_{PI}$,
we can obtain the Gamov-type relation for the spectrum of photon energy
$E_{P\gamma}=M_{P0}=m_{h0}/2,M_{W0},M_{Z0}$
in the processes $h\to \gamma\gamma$, $W^+W^-\to
\gamma\gamma$, and $ZZ \to \gamma\gamma$, respectively,
with the peaks in the CMB spectrum
 \be\label{sp-1}
 M_{P0}=k_{\rm eff}\, \frac{a_{Pd}}{a_{PI}}=k_{\rm eff}\, \ell^{1/3}_{Pd},
 \ee
where $k_{\rm eff}\simeq 9.8 \mbox{\rm GeV}$ is defined by the boson masses:
$\ell^{1/3}_{Wd}={M_{W0}}/{k_{\rm eff}}$.
The empiric formula~(\ref{sp-1}) ${M_{P0}=M_{W,Z}}=k_{\rm eff}\ell^{1/3}_{Pd}$ describes
the ratio of $W$ and $Z$ masses (\textit{i.e.} the Weinberg angle)
\be
\frac{M_Z}{M_W} = 1.134 \approx \left(\frac{800}{546}\right)^{1/3}= 1.136 ~\to ~ [\,\sin^2\theta_W \approx 0.225].
\ee
The value of Higgs particle mass is estimated as
\be\label{118}
\frac{m_h}{2}=M_W\left(\frac{\ell_1}{\ell_2}\right)^{1/3}=M_W\left(\frac{220}{546}\right)^{1/3}
\simeq 59 \,\mbox{\rm GeV},
\ee
if one takes into account that in the process $h\to 2\gamma$ the photon energy
is the half of the Higgs boson mass.
This value of the Higgs boson mass
\be\label{118-2}
m_h=118 \mbox{\rm GeV}
\ee
 is close to the present fit of the LEP
experimental data supporting rather low values just above the experimental
limit $m_h>114.4$~GeV  \cite{hhg}.


\section{Summary}
   We describe  the CMB power spectrum in accord with the well-established Wigner classification of relativistic
states treating
 General Relativity  as the theory   of  the dynamical scale symmetry
with the Poincar\'e group of the vacuum stability \cite{og}.
In particular, the CMB moving with the velocity 368 km/s to Leo
is considered as an object of Poincar\'e group transformations in order to pass in
the CMB comoving frame (\ref{dg-2}).
In this frame, the cosmological dynamics can be introduced by help of Einstein's cosmological principle (\ref{1-3})
that means averaging all scalar characteristics including the dilaton field
over a constant Universe volume.
This cosmological dynamics based on the first principles of general relativity and
relativistic invariance
 gives us a possibility to describe
 the SN Ia data by
 the ordinary free kinetic motion (\ref{1-1dg-a}) of all scalar fields (dilaton, Higgs, and $Q$) with the initial data (\ref{id-2a})
and the positive energy and vacuum postulate. In this case,
the CMB arises as final decay product of the primordial vector bosons and Higgs particles created from
the vacuum in agreement with the value of the CMB temperature and baryon number density.
The CMB power spectrum can be explained by two photon decays of these  primordial particles (\ref{sp-1})
that lead to a value of
the Higgs mass (\ref{118}) about $120$ GeV
in agreement with the Weinberg angle and QED coupling constant.

{\bf Acknowledgements}

The authors are  grateful to  E.A.~Kuraev,  A.V.~Efremov, D.~Blaschke,
 V.B.~Priezzhev,  and  S.A.~Smolyansky
for interest, criticism and creative discussions.
One of us (A.A.) acknowledges the for support the INTAS grant 05-1000008-8328
and the grant of the RF President (Scientific Schools 3312.2008.2).

\renewcommand{\theequation}{A.\arabic{equation}}

\setcounter{equation}{0}

\section*{Appendix A:  Dynamics of the dilaton}

The variation of action (\ref{1-1dg}) with respect to the lapse function ${N}_{\rm d}$ leads
to the energy constraint
 \bea\label{h-c1d}
 N_{\rm d}\frac{\delta S_{\rm U}}{\delta{N}_{\rm d}}
 &=&0 ~\to~\frac{[{\partial_{0}{\langle D\rangle}]^2
 -[\partial_{0}{\langle \phi \rangle}]^2-[\partial_{0}{\langle Q\rangle}]^2}}{N_{\rm d}}
 - N_{\rm d}T_{\rm d}=0,
 \eea
 where
 \be\label{CC-3}
 {T}_{\rm d}=-\frac{\delta \widetilde{S}_{\rm U}}{\delta {N}_{\rm d}}=\frac{4}{3}{e}^{-7D/2}
 \triangle
 {e}^{-D/2}+
  \sum\limits_{J=0,2,3,4} {e}^{-JD}{\cal T}_J(\widetilde{F})
\ee
is the local energy density
as the sum of energy  densities  ${\cal T}_J=\langle{\cal T}_J\rangle+\overline{{\cal T}}_J$ in terms of conformal fields  (\ref{1-2})
repeating cosmological regimes of the rigid state $J=0$, radiation
$J=2$, mass $J=3$, curvature $J=4$, in the SC, where $\langle{\cal T}_J\rangle=H_0^2\Omega_J$.

Averaging  the energy constraint (\ref{h-c1d}) over the volume $V_0$  leads to the global constraint
 \be\label{CC-1}
 [\partial_{\tau}\langle D\rangle]^2=[\partial_{\tau}\langle \phi\rangle]^2
 +[\partial_{\tau}\langle Q\rangle]^2+\langle \sqrt{T_{\rm d}}\rangle^2
 \ee
and  determines the diffeo-invariant lapse function
 \be\label{CC-2}{\cal N}=\frac{N_{\rm d}}{N_0}=\frac{\langle \sqrt{T_{\rm d}}\rangle}{\sqrt{T_{\rm d}}}\ee
and the diffeo-invariant interval $d\tau=N_0dx^0$ through the energy density (\ref{CC-3})

 The dilaton field ${D}=\langle D\rangle+\overline{D}$ is defined  by equation
\bea\label{h-c1D}
 \frac{\delta S_{\rm U}}{\delta D}
 =0~ \to~ 2\partial^2_\tau {\langle D\rangle}&=&{\langle T_D\rangle},\\\label{CC-4}
  (\partial_0-N^l\partial_l)P_{\overline{D}}&=&T_D-{\langle T_D\rangle},
 \eea
where \be\label{1-D}{T}_{D}=-\frac{\delta S_{\rm U}}{\delta D}=\frac{2}{3}
   \left\{7{\cal N}{e}^{7D/2}
  \triangle {e}^{D/2}+{e}^{D/2} \triangle
\left[{\cal N}{e}^{7D/2}\right]\right\}+
  {\cal N}\sum\limits_{J=0,2,3,4}J {e}^{JD}{\cal T}_J,\ee
and
$
P_{\overline{D}}=2v_{{\overline{D}}}=2\left[(\partial_0-N^l\partial_l){\overline{D}}
 +\partial_lN^l/3\right]/N_{\rm d}
 $ 
is the dilaton momentum.

Eqs. (\ref{CC-1}),~(\ref{h-c1D}) can be treated as the exact analogy of the Friedmann equation
in the tangent space-time defined by simplex components
\be\label{1-Dn}
\tilde \omega _{(0)}  = e^{ - 2 D} {\cal N}d\tau
= e^{ - 2 \overline{D}} \frac{{\left\langle {\sqrt {T_d } } \right\rangle }}{{\sqrt {T_d } }}d\eta,
\qquad \tilde \omega _{(b)}={\bf e}_{(b)k}dx^k+{\cal N}_{(b)}d\eta.
\ee

The Hamiltonian approach to the theory (\ref{1-1dg}) was considered in \cite{Barbashov_06} in the Dirac
gauge~\cite{dir}
 $
  P_{\overline{D}}=0, ~\partial_k {\bf e}^k_{(b)}=0
  $.


\renewcommand{\theequation}{B.\arabic{equation}}

\setcounter{equation}{0}

\section*{Appendix B: The Newton law status in cosmological perturbation theory}

The investigation of the large-scale structure in the Early Universe is one of the highlights of
the present-day Cosmology with far-reaching implications.
 In particular, the comparison of the cosmological perturbation theory in  the $\Lambda$CDM model
 with the Hamiltonian approach to the same cosmological perturbation theory \cite{Barbashov_06}
 reveals essential differences of these approaches and their physical consequences.

 In order to demonstrate these consequences, we consider the case of integrable
  diffeo-invariant spacial coordinates, when the simplex components in interval (\ref{dg-2})
  ${\bf e}_{(b)i}dx^i={\omega}^{(3)}_{(b)}=dx_{(b)}$
 are total differentials.  The latter means
  that the coefficients of the spin-connection are equal to zero
  $
  \sigma_{(a)|(b)(c)}=
 {\bf e}_{(a)j}\left[\partial_{(b)}{\bf e}^j_{(c)}
 -\partial_{(c)}{\bf e}^j_{(b)}\right]=0
  $
  together with the  three-dimensional curvature $R^{(3)}=0$ in accord with observational data \cite{WMAP08}. 
%
 In this case, the transverse components of the shift vector can be defined by
 \bea\label{h-c1s}
 T_{(0)(a)}&=&-{\bf e}_{(b)}^i\frac{\delta S_{\rm U}}
{\delta {N}_{i}}=
  -\partial_{(b)}p_{(b)(a)}+
  \sum\limits_{f=\overline{\phi},\overline{Q},\widetilde{F}}^{}p_f\partial_{(a)}f=0,
\\\label{ls-1}
   p_{(b)(a)}&=&\frac{1}{3}v_{(ab)}=\frac{1}{6{\cal N}}\left(\frac{2}{3}\delta_{(a)(b)}\partial_{(c)}{\cal N}_{(c)}-\partial_{(a)}{\cal N}_{(b)}-\partial_{(b)}{\cal N}_{(a)}\right).
  \eea
 While the shift vector longitudinal component   is given by the Dirac constraint
 $
 \partial_\eta e^{-3\overline{D}}=
 \partial_{(b)}\left(e^{-3\overline{D}}{\cal N}_{(b)}\right)
 $. 
 The lapse function and dilaton are determined as solutions of Eqs.~(\ref{CC-2}),~and~(\ref{CC-4}).
  Solutions of these Eqs., in the first order in the Newton coupling constant,
  take forms \cite{Barbashov_06,B_06}
  \bea\label{12-17}
 e^{-\overline{D}/2}
 &=&1+\frac{1}{2}\int d^3y\left[G_{(+)}(x,y)
\overline{T}_{(+)}^{(\mu)}(y)+
 G_{(-)}(x,y) \overline{T}^{(\mu)}_{(-)}(y)\right],\\\label{12-18}
 {\cal N}e^{-7\overline{D}/2}
 &=&1-\frac{1}{2}\int d^3y\left[G_{(+)}(x,y)
\overline{T}^{(\nu)}_{(+)}(y)+
 G_{(-)}(x,y) \overline{T}^{(\nu)}_{(-)}(y)\right],
  \eea
 where
$D_{(\pm)}(x,y)$ are the Green functions satisfying
 the equations
 \bea\label{2-19}
 &&[\pm  m^2_{(\pm)}- \triangle
 ]G_{(\pm)}(x,y)=\delta^3(x-y),\\\label{2-19-1}
 &&m^2_{(\pm)}= 
 \frac{3(1+z)^2}{4}\left[{14(\beta\pm
1)}\Omega_{(0)}(z) \!\mp\!
 \Omega_{(1)}(z)\right]H_0^2,\\\label{2-19-2}
 &&\beta=\sqrt{\!1+\![\Omega_{(2)}(z)\!-\!\!14\Omega_{(1)}(z)
 ]/[98\Omega_{(0)}(z)]},\\\label{2-19-3}
 &&\Omega_{(n)}(z)=\sum\limits_{J=0,2,3,4,6}J^n(1+z)^{2-J}\Omega_{J}, \quad
 \Omega_{J} = \langle{\cal T}_J\rangle/H_0^2.
 \eea
$\Omega_{J=0,2,3,4,6}$ are partial  density of states: rigid,
radiation, matter, curvature, $\Lambda$-term, respectively; $\Omega_{(0)}(0)=1$,
and $H_0$ is Hubble parameter,
 \bea\label{1cur1}\overline{T}^{(\mu)}_{(\pm)}
 &=&\overline{{\cal T}}_{(0)}\mp7\beta
  [7\overline{{\cal T}}_{(0)}-\overline{{\cal T}}_{(1)}],\\
  ~\overline{T}^{(\nu)}_{(\pm)}&=&[7\overline{\cal T}_{(0)}
 -\overline{\cal T}_{(1)}]
 \pm(14\beta)^{-1}\overline{\cal T}_{(0)}
 \eea
 are the local currents.

In the case of point mass distribution in a finite volume $V_0$ with the
zeroth pressure
  and  the  density
 \be\overline{{\cal T}}_{(0)}(x)=\frac{\overline{{\cal T}}_{(1)}(x)}{6}
  \equiv \frac{3}{4a^2} M\left[\delta^3(x-y)-\frac{1}{V_0}\right],\ee
 solutions   (\ref{12-17}),  (\ref{12-18}) take
 the Schwarzschild -type   form
 \bea\nonumber
  e^{-\overline{D}/2}&=&1+
  \frac{r_{g}}{4r}\left[\frac{1+7\beta}{2}e^{-m_{(+)}(z)
 r}+ \frac{1-7\beta}{2}\cos{m_{(-)}(z)
 r}\right]_{H_0=0}=1+
  \frac{r_{g}}{4r},
 \\\nonumber
 {\cal N}e^{-7\overline{D}/2}&=&1-
 \frac{r_{g}}{4r}\left[\frac{14\beta+1}{28\beta}e^{-m_{(+)}(z)
 r}+ \frac{14\beta-1}{28\beta}\cos{m_{(-)}(z)
 r}\right]_{H_0=0}=1-
 \frac{r_{g}}{4r},
 \eea
where $\beta=\sqrt{{25}/{49}}\simeq 1.01/\sqrt{2}$,~
$m_{(+)}=3m_{(-)}, m_{(-)}=H_0\sqrt{(1+z)\Omega_M \,3/2}
$.
These solutions have spatial oscillations and the
nonzero shift of the coordinate
  origin.

One can see that in the infinite volume limit $H_0=0,~a=1$
 these solutions  coincide with
 the isotropic version of  the Schwarzschild solutions:
 $e^{-\overline{D}/2}=1+\frac{r_g}{4r}$,~
 ${\cal N}e^{-7\overline{D}/2}=1-\frac{r_g}{4r}$,~$N^k=0$.
 However,
any nonzero cosmological density $\langle{T}^{1/2}_{\rm d}\rangle>0$ forbids negative values of the lapse function 
 ${\cal N}= {\langle{T}^{1/2}_{\rm d}\rangle}/
{{T}^{1/2}_{\rm d}}>0$ \cite{Barbashov_06,B_06,Glinka_08}.

%
%

\renewcommand{\theequation}{C.\arabic{equation}}

\setcounter{equation}{0}
\section*{Appendix C: Baryon-antibaryon asymmetry}

In SM, in each of  the three generations of leptons
 (e,$\mu$,$\tau$) and color quarks, we have four fermion
 doublets -- in all there are $n_L=12$ of them. Each of 12 fermion
 doublets interacts with the triplet of non-Abelian fields
 $A^1=(W^{(-)}+W^{(+)})/\sqrt{2}$, $A^2=
 i(W^{(-)}-W^{(+)})/\sqrt{2}$, and $A^3=Z/\cos\theta_{(W)}$,
 the corresponding coupling constant being $g=e/\sin\theta_{(W)}$.
 It is well known that, because of a triangle anomaly, W- and
Z- boson interaction with lefthanded fermion doublets
 $\psi_L^{(i)}$, $i=1,2,...,n_L$, leads to
 nonconservation of the number of fermions of each
type ${(i)}$
\cite{ufn},
 \bea \label{rub}
 \partial_\mu j^{(i)}_{L\mu}=\frac{1}{32\pi^2}
 {\rm Tr}\hat F_{\mu\nu}{}^*\!{\hat F_{\mu\nu}},
 \eea
 where $\hat
 F_{\mu\nu}=-iF^a_{\mu\nu}g_W\tau_a/2$ is the strength of the
 vector fields, $F^a_{\mu\nu}=
 \partial_\mu A_\nu^a-\partial_\nu
 A_\mu^a+g\epsilon^{abc}A_\mu^bA_\nu^c$.

 Taking the integral of the equality in (\ref{rub}) with respect to conformal time and
 the three-dimensional variable $x$, we can find a relation between
 the change
 \be \label{rub-1}
 \int\limits_{\eta_I}^{\eta_0} d\eta \int d^3x
 \partial_\mu j^{(i)}_{L\mu}=F^{(i)}(\eta_0)-F^{(i)}(\eta_I) =\Delta F^{(i)}\ee
  of the fermion number $ F^{(i)}=\int d^3x
 j_0^{(i)}$ and the Chern-Simons functional
 $ 
 F_{\mu\nu}{}^*\!{\hat F_{\mu\nu}},
 $ 
 so that after integration Eq.  (\ref{rub}) takes the form
  \be \label{rub2}
 \Delta F^{(i)}= N_{CS} \not = 0, ~~~i=1,2,...,n_L.
 \ee
 The equality in (\ref{rub2}) is considered as a selection rule --
that is, the fermion number changes identically for all fermion
types:
 $N_{CS}=\Delta L^e=\Delta L^\mu=\Delta L^\tau=\Delta B/3$;
 at the same time, the change in the baryon charge $B$ and the change
 in the lepton charge  $L=L^e+L^\mu+L^\tau$ are related to each other in
such a way that $B-L$ is conserved, while
 $B+L$ is not invariant. Upon taking the sum of the equalities in
 (\ref{rub2}) over all doublets, one can  obtain $\Delta (B+ L)=12
 N_{CS}$ ~\cite{ufn}.

 We can evaluate the expectation value of the Chern-Simons
 functional (\ref{rub2})  (in the lowest order of perturbation
 theory in the coupling constant) in the Bogoliubov vacuum
 $b|0>=0$. Specifically, we have
 \be
 N_{CS}=N_{\rm W}\equiv
 -\frac{1}{32\pi^2}\int_0^{\eta_{L}} d\eta \int {d^3 x} \;
 \langle 0|{\rm Tr}\hat F^{\rm W}_{\mu\nu}
 {}^*\!{\hat F^{\rm W}_{\mu\nu}}|0\rangle ,
 \ee
 where $\eta_{L}$  is the W-boson
 lifetime, and $N_{\rm W}$
 is the contribution
 of the primordial $W$  boson.
 The integral over the conformal spacetime bounded
 by three-dimensional hypersurfaces $\eta=0$ and $\eta =\eta_L$
  is given by
 $N_{\rm W} ={2}{\alpha_W}V_0
 \int_{0}^{{\eta_{L}}} d\eta \int\limits_{0 }^{\infty }dk
 |k|^3 R_{\rm W}(k,\eta)
 $,
 where
 ${\alpha_W}={{\alpha}_{\rm
 QED}}/{\sin^{2}\theta_{W}}$
 and
 $R_{\rm W}=\frac{i}{2}{}_b<0|b^+b^+-b^-b^-|0>_b=-\sinh(2r(\eta_L))\sin(2\theta(\eta_L))
 $
  is the Bogoliubov condensate \cite{Blaschke_04} that is
 specified by relevant solutions to the
 Bogoliubov equations. Upon a numerical calculation
 of this integral, we can estimate the expectation value of the
 Chern-Simons functional in the state of primordial bosons.

 At the vector-boson-lifetime value in (\ref{lv}),
 this yields the following result at 
 $n_{\gamma}={ 2,402  \times T^3 }/{\pi^2}$
\be
\frac{N_{CS}}{V_{0}}=
\frac{ N_W}{V_{0}}
 = 4\alpha_W 
  T^3
 \times 1.44
  =0.8~  n_{\gamma}.
\ee
where $n_{\gamma}$ is the number
 density of photons forming Cosmic Microwave Background radiation.
On this basis, the violation of the fermion-number density in
the cosmological model being considered can be estimated as
\cite{Blaschke_04,Behnke_02}
$
{\Delta F^{(i)}}/{V_{0}}={N_{CS}}/{V_{0}}
  =0.8  n_{\gamma}
$. 

 According to SM, there is the CKM-mixing
 that leads to ${\rm CP}$ nonconservation,
 so that the cosmological evolution and
 this  nonconservation freeze
  the
  fermion number at $\eta=\eta_L$.  This
 leads to the baryon-number density \cite{ufn,sufn}
$ 
  n_{\rm b}(\eta_L)=
  X_{\rm CP}{\Delta
  F^{(i)}}/{V_{0}}\simeq X_{\rm CP}n_{\gamma}(\eta_L)
$,  
  where the factor $X_{\rm CP}$ is determined by the superweak
 interaction of $d$ and $s$ quarks,
 which
 is responsible for CP violation experimentally observed in
 $K$-meson decays \cite{o}.

 From the ratio of the number of baryons to the number of photons,
 one can deduce an estimate of the superweak-interaction coupling
 constant: $X_{\rm CP}\sim 10^{-9}$.
  Thus, the evolution of the Universe, primary
 vector bosons, and the aforementioned superweak interaction~\cite{o}
  lead to baryon-antibaryon
 asymmetry of the Universe
 \be\label{data6a} \frac{n_{\rm b}(\eta_L)}{n_{\gamma}(\eta_L)}\simeq X_{\rm CP}= 10^{-9}.
  \ee


 Thus, the primordial bosons before
 their decays polarize the Dirac fermion vacuum and give the
 baryon asymmetry frozen by the CP -- violation
 so that for billion photons there is only one baryon.

The problem is to show that the Universe matter content
considered as the final decay product of primordial bosons is in agreement with
observational data~\cite{Blaschke_04}.

\end{document}